\documentclass[citeautoscript,floatfix,superscriptaddress,twocolumn,showpacs,prb,aps,10pt]{revtex4}

\usepackage{graphicx} \usepackage{amsmath, amssymb, bm}
\usepackage{dcolumn} \usepackage{color}
\usepackage{eso-pic}
\usepackage{fix-cm}

\usepackage{multirow}
\usepackage{sidecap}

\usepackage{marginnote}
\usepackage{ulem} 

\renewcommand{\emph}[1]{\textit{#1}}

\begin{document}
\title{Elevated temperature dependence of the anisotropic visible-to-ultraviolet dielectric function of monoclinic $\beta$-Ga$_2$O$_3$}

\author{A. Mock}
\email{amock@huskers.unl.edu}
\homepage{http://ellipsometry.unl.edu}
\affiliation{Department of Electrical and Computer Engineering and Center for Nanohybrid Functional Materials, University of Nebraska-Lincoln, U.S.A.}
\author{J. VanDerslice}
\affiliation{J. A. Woollam Co., Inc., 645 M St. Lincoln NE, USA}
\author{R.~Korlacki}
\affiliation{Department of Electrical and Computer Engineering and Center for Nanohybrid Functional Materials, University of Nebraska-Lincoln, U.S.A.}
\author{J. A. Woollam}
\affiliation{Department of Electrical and Computer Engineering and Center for Nanohybrid Functional Materials, University of Nebraska-Lincoln, U.S.A.}
\affiliation{J. A. Woollam Co., Inc., 645 M St. Lincoln NE, USA}
\author{M. Schubert}
\affiliation{Department of Electrical and Computer Engineering and Center for Nanohybrid Functional Materials, University of Nebraska-Lincoln, U.S.A.}
\affiliation{Leibniz Institute for Polymer Research, Dresden, Germany}
\affiliation{Department of Physics, Chemistry, and Biology, Link{\"o}ping University, 58183 Link{\"o}ping, Sweden}

\date{}

\begin{abstract}

We report on the temperature dependence of the dielectric tensor elements of $n$-type conductive $\beta$-Ga$_2$O$_3$ from 22$^\circ$C-500$^\circ$C in the spectral range of 1.5~eV--6.4~eV. We present the temperature dependence of the excitonic and band-to-band transition energies and their eigenpolarization vector orientations. We utilize a Bose-Einstein analysis of the temperature dependence of the observed transition energies and reveal electron coupling with average phonon temperature in excellent agreement with  the average over all longitudinal phonon plasmon coupled modes reported previously [M. Schubert~\textit{et al.}, Phys. Rev. B \textbf{93}, 125209 (2016)]. We also report a linear temperature dependence of the wavelength independent Cauchy expansion coefficient for the anisotropic below-band-gap monoclinic indices of refraction.


\end{abstract}
 \maketitle

Recently, the ultra-wide band gap semiconductor with monoclinic crystal symmetry, $\beta$-Ga$_2$O$_3$, has become the subject of much research due to its potential for applications in transparent electronics and high-energy photonics but also due to its potential to replace GaN and SiC in next generation power electronics. The monoclinic $\beta$-phase of gallium oxide has the widest range of thermodynamic stability among the five polytypes ($\alpha,~\beta,~\gamma,~\delta,~\textrm{and}~\epsilon$).\cite{Roy_1952,Tippins_1965} Knowledge about fundamental properties of semiconductors with monoclinic symmetry is not exhaustive. $\beta$-Ga$_2$O$_3$ is attracting attention because of high electrical breakdown field of 8~MV/cm and room-temperature electron mobility of 300~cm$^2$/Vs.\cite{Higashiwaki_2012,Higashiwaki_2013, Green_2016,Wong_2016} Schottky devices with reverse breakdown voltage in excess of 1~kV were reported.\cite{Yang_2017,Konishi_2017} $\beta$-Ga$_2$O$_3$ has an ultra-wide room-temperature band-gap of 5.04~eV,\cite{Mock_2017Ga2O3} which has been the topic of expanding computational and experimental work.\cite{Wager_2003,Sturm_2015,Sturm_2016,Mock_2017Ga2O3,Furthmuller_2016} Ultra-violet solar-blind photo detectors have been reported.\cite{Kokubun_2007,Oshima_2008,Suzuki_2011} Transparent devices with $\beta$-Ga$_2$O$_3$ may operate at elevated temperatures, and the temperature dependence of the properties of this emerging semiconductor are of interest. Recent combined investigations with generalized spectroscopic ellipsometry (GSE) and density functional theory into the optical properties of this monoclinic semiconductor were made to explore the room temperature dielectric tensor element spectra, and to identify electronic and excitonic properties in single crystalline $\beta$-Ga$_2$O$_3$.\cite{Mock_2017Ga2O3} Excitonic contributions were found to have distinct binding energies for different band pairs, unlike zincblende or wurtzite structure semiconductors, as a consequence of the highly anisotropic monoclinic lattice system. Also recently, the complete infrared active phonon mode properties were revealed by DFT and GSE investigations.\cite{Schubert_2016} In particular, the coupling behavior of longitudinal optical (LO) modes with plasmon modes in $n$-type doped $\beta$-Ga$_2$O$_3$ was described and found to differ fundamentally from that of traditional semiconductor materials such as Si and GaAs. Longitudinal plasmon-phonon (LPP) modes were found to differ in their polarization directions from each other, and a strong dependence on the eigendielectric polarization directions on the charge carrier density was predicted. In a very recent paper, Sturm~\textit{et al.} report on the evolution of the dielectric function tensor in the visible-to-ultra-violet spectral regions from room temperature towards low temperatures. The dependence of the exciton and band-to band transition was analyzed using the Bose-Einstein model.\cite{Sturm_2017} In this work we investigate the effect of elevated temperature onto the dielectric function of $n$-type doped $\beta$-Ga$_2$O$_3$ and derive therefrom excitonic and band-to-band transition energies using an eigendielectric polarization model approach.\cite{Mock_2017Ga2O3} In the eigendielectric polarization approach, critical point structures which contribute to the anisotropic dielectric function tensor of $\beta$-Ga$_2$O$_3$ are represented by direction dependent (dyadic) polarizability functions.\cite{SchubertPRL2016} The direction dependence originates from the polarization selection rules of the interband matrix elements.\cite{Mock_2017Ga2O3} For monoclinic symmetry materials, an interesting question is whether and under what circumstance not only exciton and band-to-band transitions shift but whether their polarization selection rules change.

\begin{figure*}[hbt]
\centering
\includegraphics[width=\linewidth]{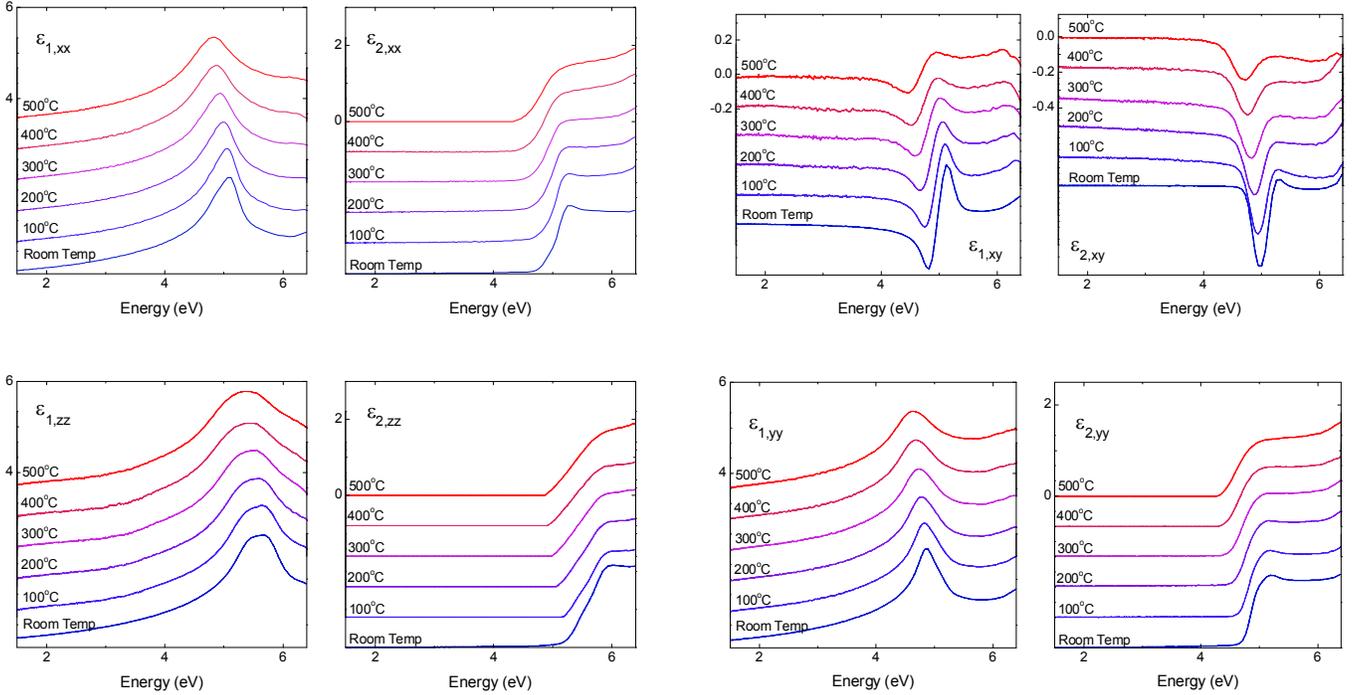}
\caption{\label{fig:dielectric} Real and imaginary components of the dielectric tensor elements, $\varepsilon_{xx}$, $\varepsilon_{yy}$, $\varepsilon_{xy}$, and $\varepsilon_{zz}$ for room temperature (blue) to 500$^\circ$C (red). The functions were shifted vertically with respect to the function at room temperature for convenience.}
\end{figure*}

Two single side polished crystallographic surfaces, (010) and ($\bar{2}$01), cut from a single crystal of Ga$_2$O$_3$  grown by Tamura Corp., Japan by edge-defined film fed growth process\cite{Aida_2008,Sasaki_2013,Shimamura_2013} were investigated. Mueller matrix spectroscopic ellipsometry data were collected in the spectral range of 194--1660~nm using a dual-rotating compensator ellipsometer (RC2, J.~A.~Woollam~Co.,~Inc.). Samples were placed inside a nitrogen purged heating cell (Heat Cell, J.~A.~Woollam~Co.,~Inc.) and aligned at an angle of incidence of 70$^\circ$. Data was acquired \textit{in-situ} while the sample chuck was heated from room temperature to 575$^\circ$C at a constant rate of 6~$^\circ$C/s. Temperature dependent Mueller matrix data was collected from 3 azimuthal orientations per sample, by manually rotating each sample clockwise by sample normal in steps of approximately 45$^\circ$. We note that window effects on the Mueller matrix elements were accounted for by utilizing a reference bulk Si wafer. All model calculations reported in this work were performed using WVASE32$^{TM}$ (J.~A.~Woollam~Co.,~Inc.)

We choose a coordinate system as described in Ref.~ \onlinecite{Mock_2017Ga2O3} with $x$ parallel to \textbf{a}, $y$ parallel to \textbf{c$^{\star}$}, and $z$ parallel to \textbf{b} yielding the dielectric tensor:
\begin{equation}
\varepsilon = \begin{pmatrix} \varepsilon_{xx} & \varepsilon_{xy} & 0\\ \varepsilon_{xy} & \varepsilon_{yy} & 0\\ 0 & 0 & \varepsilon_{zz}\end{pmatrix}.
\end{equation}
Note that \textbf{c$^{\star}$} is defined perpendicular to the \textbf{a}--\textbf{b} plane for convenience, while \textbf{c} shares the monoclinic angle $\beta$=103.7$^{\circ}$ with  \textbf{a}.\cite{Geller_1960}  A wavelength-by-wavelength approach is utilized at each temperature simultaneously fit to 6 independent data sets (2 samples with 3 orientations each) to obtain the dielectric tensor elements $\varepsilon_{xx}$, $\varepsilon_{yy}$, $\varepsilon_{xy}$, and $\varepsilon_{zz}$. We note that surface roughness was accounted for as described in Ref.~ \onlinecite{Mock_2017Ga2O3}. Oscillators are then projected into each direction as well as into the shear plane. From the parameters of these oscillator functions, amplitude, broadening, excitonic energy, transition energies, and their eigendielectric polarization orientation are determined. Details pertaining to this model approach can be found in Ref.~ \onlinecite{Mock_2017Ga2O3}



\begin{figure}
\centering
\includegraphics[width=\linewidth]{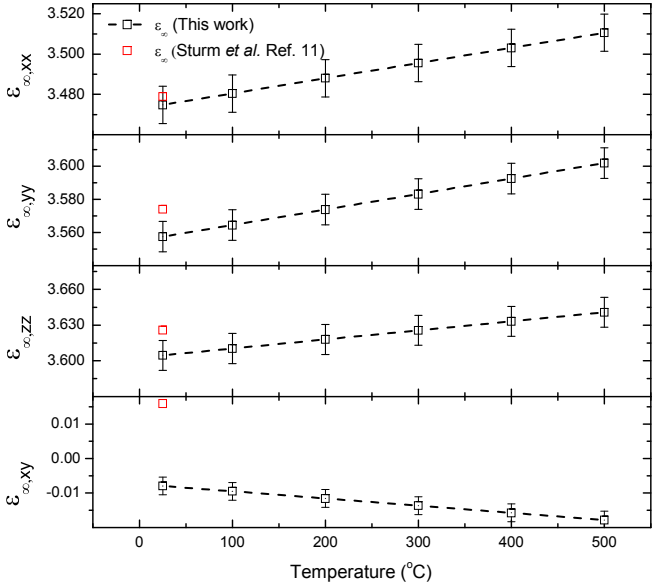}
\caption{\label{fig:cauchy} Wavelength independent part of the square of the anisotropic indices of refraction of monoclinic $\beta$-Ga$_2$O$_3$ determine from the below-band-gap dielectric function tensor as a function of temperature (black squares), and the best-match linear interpolation (dashed lines).}
\end{figure}

Figure~\ref{fig:dielectric} depicts the real and imaginary parts of the dielectric function obtained at temperatures 22$^\circ$C, 100$^\circ$C, 200$^\circ$C, 300$^\circ$C, 400$^\circ$C, and 500$^\circ$C and determined by a wavelength-by-wavelength approach. We observe a distinct broadening accompanied with a pronounced red-shift of the critical point features. To begin with, the below-band-gap squares of the indices of refraction are analyzed, determining $\varepsilon_\infty$ for each non-vanishing tensor element. A Cauchy polynomial expansion was used retaining the first three coefficients in the wavelength expansion, augmented by a linear dependence on temperature for the wavelength independent Cauchy coefficient to analyze the wavelength-by-wavelength determined dielectric tensor elements:
\begin{equation}
\varepsilon_{1,ij} = \varepsilon_{\infty,ij}+A_{ij}(\textrm{T}-\textrm{T}_0)+\frac{B_{ij}}{\lambda^2}+\frac{C_{ij}}{\lambda^4},
\label{eq:Cauchy}
\end{equation}
where T$_0$ is room temperature. Figure~\ref{fig:cauchy} presents the the best-match model parameter result for the wavelength independent portion of Eq.~\ref{eq:Cauchy}, obtained for each temperature investigated (black, boxes), and as a function of temperature (dashed lines). Table~\ref{tab:Linearparms} lists all below-band-gap model parameters, and compares $\varepsilon_{\infty}$ values reported at room temperature previously (Sturm~\textit{et al.},~Ref.~\onlinecite{Sturm_2015}), where we note reasonable agreement except in the shear element $\varepsilon_{xy}$.

\begin{table}
\caption{\label{tab:Linearparms} Best-match model calculated below-band-gap square of the index of refraction wavelength and temperature dependence parameter values according to Eq.~\ref{eq:Cauchy} for monoclinic $\beta$-Ga$_2$O$_3$. Digits in parenthesis denote the 90\% confidence interval determined using the numerical uncertainties determined for each data point during the wavelength-by-wavelength dielectric function tensor analysis.}
\begin{tabular}{{l}{c}{c}{c}{c}}
\hline \hline
& $\varepsilon_{1,xx}$ & $\varepsilon_{1,yy}$ & $\varepsilon_{1,zz}$ & $\varepsilon_{1,xy}$ \\

\hline
$\varepsilon_{\infty,ij}$&3.4(7)&3.5(6)&3.6(0)&-0.00(8)\\
$A$ ($10^{-5}\textrm{C}^{-1}$) &(8)&(9)&(8)&-2.(0)\\
$B$ ($10^{-2}\mu\textrm{m}^2$) &(5)&(6)&(6)&0.(4)\\
$C$ ($10^{-3}\mu\textrm{m}^4$) &(6)&(4)&-(2)&-(1)\\
\hline\
$\varepsilon_{\infty,ij}$&3.7(5)$^a$&3.2(1)$^a$&3.7(1)$^a$&-0.0(8)$^a$\\
$\varepsilon_{\infty,ij}$&3.479$^b$&3.574$^b$&3.626$^b$&0.016$^b$\\
\hline \hline
\end{tabular}
\begin{tabular}{{l}}
$^{a}$ FIR-IR-GSE phonon analysis, Schubert~\textit{et al.}, Ref.~\onlinecite{Schubert_2016}.\\
$^{b}$ Room temperature VIS-VUV-GSE analysis, Sturm~\textit{et al.}, Ref.~\onlinecite{Sturm_2015}.\\
\end{tabular}
\end{table}

An eigendielectric displacement vector approach as developed by us previously\cite{Schubert_2016} has been utilized to describe electronic transitions and excitonic effects of $\beta$-Ga$_2$O$_3$ at room temperature\cite{Mock_2017Ga2O3}. We use this same approach here at elevated temperatures and present model dielectric function parameters in Tab. \ref{Tab:parms}.  We find that amplitudes do not change significantly in the temperature range investigated and are therefore not included in the table. We observe that energy parameters decrease with increasing temperature while broadening parameters increase. We note that due to increasing broadening and with the first two transitions polarized along the b-axis so close together, sensitivity to the excitonic binding energy parameter was limited and thus we held it to a constant value of 0.18~eV for CP$^{b}_{0x}$ as determined previously for room temperature.

\begin{table*}
\caption{\label{Tab:parms} Parameter values determined from the eigendielectric displacement vector approach for the temperatures investigated in this work. Digits in parenthesis denote 90\% confidence determined from this analysis. Note amplitude parameters did not change significantly from those presented previously\cite{Mock_2017Ga2O3}, thus are not presented here.}
\begin{tabular}{{l}{c}{c}{c}{c}{c}{c}{c}{c}{c}{c}{c}{c}{c}{c}}
\hline \hline
             & \multicolumn{3}{c}{CP$^{ac}_{0x}$} & \multicolumn{2}{c}{CP$^{ac}_{0}$}&\multicolumn{3}{c}{CP$^{ac}_{1x}$}&\multicolumn{2}{c}{CP$^{ac}_{1}$}&\multicolumn{2}{c}{CP$^{b}_{0x}$}&\multicolumn{2}{c}{CP$^{b}_{0}$}\\
Temperature&\multicolumn{1}{c}{$\alpha$ (deg)}&\multicolumn{1}{c}{$E$(eV)}&$B$(eV)&\multicolumn{1}{c}{$E$(eV)}&$B$(eV)&\multicolumn{1}{c}{$\alpha$ (deg)}&\multicolumn{1}{c}{$E$(eV)}&$B$(eV)&\multicolumn{1}{c}{$E$(eV)}&$B$(eV)&\multicolumn{1}{c}{$E$(eV)}&$B$(eV)&\multicolumn{1}{c}{$E$(eV)}&$B$(eV)\\
\hline
RoomTemp&115.1(1)&4.9(2)&0.4(0)&5.0(4)&0.0(2)&25.2(1)&5.1(7)&0.4(3)&5.4(0)&0.0(9)&5.4(6)&0.5(4)&5.6(4)&.01(1)\\
100$^\circ$C&112.(5)&4.8(7)&0.46(6)&5.0(1)&0.03(2)&25.(2)&5.1(4)&0.45(0)&5.3(1)&0.08(1)&5.41(7)&0.5(3)&5.59(6)&1.(4)\\
200$^\circ$C&111.(7)&4.8(3)&0.5(2)&4.9(6)&0.0(6)&20.(3)&5.0(8)&0.53(8)&5.2(9)&0.0(9)&5.34(9)&0.6(1)&5.52(9)&1.(6))\\
300$^\circ$C&111.(7)&4.7(6)&0.5(9)&4.9(2)&0.0(7)&20.(1)&5.0(2)&0.5(9)&5.2(2)&0.1(3)&5.3(0)&0.6(8)&5.4(8)&2.(0)\\
400$^\circ$C&110.(4)&4.6(9)&0.6(2)&4.8(5)&0.0(8)&19.(8)&4.9(3)&0.6(9)&5.0(9)&0.(5)&5.2(8)&0.8(1)&5.4(6)&3.(2)\\
500$^\circ$C&109.(6)&4.6(2)&0.6(7)&4.7(9)&0.0(9)&18.(0)&4.8(6)&0.8(0)&5.1(3)&0.(7)&5.2(5)&0.8(9)&5.4(3)&3.1(9)\\
\hline \hline
\end{tabular}
\end{table*}

\begin{figure}
\centering
\includegraphics[width=\linewidth]{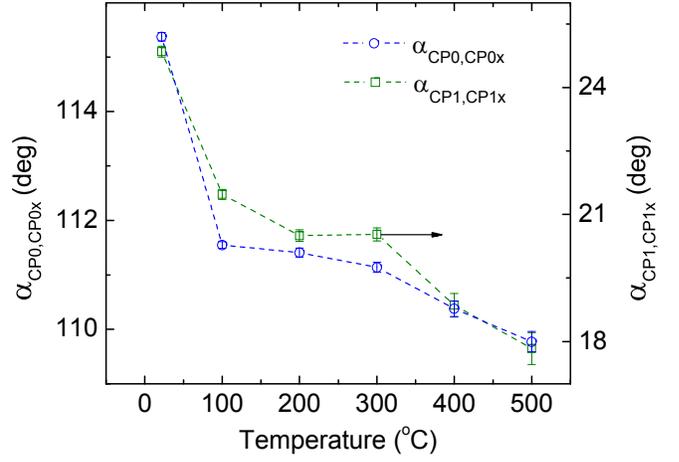}
\caption{\label{fig:directions}  Temperature dependence of polarization direction for critical points CP$^{ac}_{0x}$ and CP$^{ac}_{0x}$ (blue) as well as for CP$^{ac}_{1x}$ and CP$^{ac}_{1}$ (green). Dashed lines are provided as a guide for the eye.}
\end{figure}

The evolution of the eigendielectric displacement vector direction with temperature is presented in Fig. \ref{fig:directions}. We observe that in the case of the first two transitions within the \textbf{a-c} plane (CP$_0$ and CP$_1$) along with their excitonic contributions, both of their corresponding eigendielectric displacement vectors orientation angles ($\alpha$) decrease. Therefore, we observe that with increasing temperature the direction of the transition corresponding to CP$_0$ (and CP$_{0x}$) shift closer to the \textbf{c} axis which is at approximately 103.73$^\circ$. In much the same way, the direction of the transition corresponding to CP$_1$ (and CP$_{1x}$) shifts closer to the \textbf{a} axis.

\begin{figure}
\centering
\includegraphics[width=\linewidth]{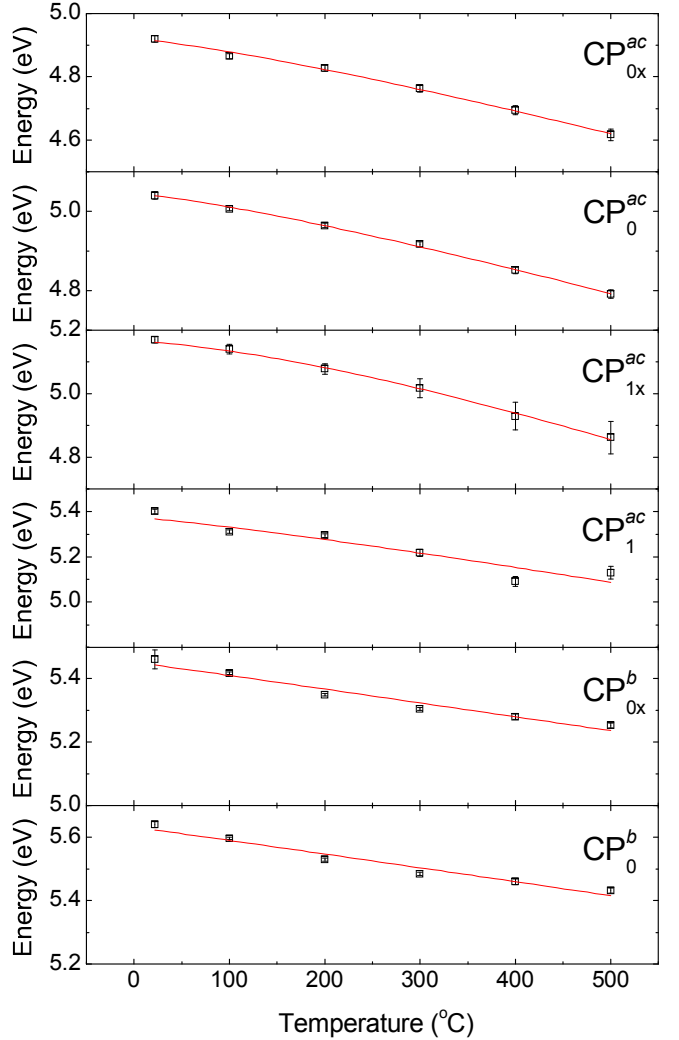}
\caption{\label{fig:bose-einstein}  Temperature dependence of energy and broadening parameters for critical points CP$^{ac}_{0x}$, CP$^{ac}_{0x}$, CP$^{ac}_{1x}$, CP$^{ac}_{1}$, CP$^{b}_{0x}$, CP$^{b}_{0}$. The red solid lines indicate the Bose-Einstein model lineshape.}
\end{figure}

\begin{table}
\caption{\label{Tab:B-Efit} Parameter values determined from a Bose-Einstein model of temperature dependent parameter values.}
\begin{tabular}{{l}{c}{c}{c}{c}{c}}
\hline \hline
CP & $E_0$ (eV) & $\alpha$ (unitless) & $P$ (eV) & $\alpha$ (Ref. \onlinecite{Sturm_2017})\footnote{Parameter average values determined for \textbf{a-c} plane and \textbf{b} axis separately by Sturm \textit{et al.}} & $P$ (Ref. \onlinecite{Sturm_2017})$^a$ \\
\hline
CP$^{ac}_{0x}$&4.949\footnote{Values approximated from Ref. \onlinecite{Sturm_2017} for zero temperature energies.}&9.(5)&0.0(8)&10.5&0.050\\
CP$^{ac}_{0}$&5.069$^b$&8.(0)&0.08(0)&10.5&0.050\\
CP$^{ac}_{1x}$&5.18$^b$&1(3)&0.1(1)&10.5&0.050\\
CP$^{ac}_{1}$&5.41$^b$&(8)&0.0(7)&10.5&0.050\\
CP$^{b}_{0x}$&5.52$^b$&5.(2)&0.0(3)&5.5&0.025\\
CP$^{b}_{0}$&5.7$^b$&(5)&0.0(2)&5.5&0.025\\
\hline \hline
\end{tabular}
\end{table}

Temperature dependent energy parameters for critical points are shown in Fig. \ref{fig:bose-einstein}. The Bose-Einstein model as described by Vi\~na \textit{et al.} (Ref. \onlinecite{Vina_1979}) can be used to describe temperature dependence effects to render the shifts in energy parameters according to
\begin{equation}\label{eq:bose-einstein_energy}
E(T) = E_0-\frac{\alpha P}{exp(P/k_BT)-1},
\end{equation}
where $\alpha$ is the phonon interaction strength and P is the phonon frequency as determined by temperature dependent energy shifts. Recently, Sturm \textit{et al.} (Ref. \onlinecite{Sturm_2017}) applied this technique for low temperature generalized ellipsometry analysis of single crystalline $\beta$-Ga$_2$O$_3$. We also apply this model and extend it to our high temperature measurements.

The resulting lineshapes are shown as red solid lines in Fig. \ref{fig:bose-einstein} and resulting parameters are given in Tab. \ref{Tab:B-Efit}. We note that in the temperature range investigated, there is limited sensitivity to the parameter $E_0$, therefore, we have fixed this parameter to a value estimated from results published by Sturm \textit{et al.} (Ref.~ \onlinecite{Sturm_2017}).  We find an average isotropic phonon frequency to be approximately 0.65~eV (524.26~cm$^{-1}$) from all energy parameters for the transitions investigated. The arithmetic average of all (symmetry independent) LPP phonons found by Schubert \textit{et al.} was found to be 507.5~cm$^{-1}$ which is in excellent agreement with the value found in the present study.

In summary, we present the dielectric function tensor elements of $\beta$-Ga$_2$O$_3$ from room temperature up to 500$^\circ$C. Further, we provide a description of the linear temperature dependence of the wavelength independent Cauchy coefficient to obtain the temperature effects on the anisotropic below-band-gap monoclinic indices of refraction. Additionally, we determine shifts in eigendielectric vector displacement approach energy parameters due to changes in temperature and model these effects by utilizing a Bose-Einstein lineshape to determine average phonon frequencies. Further, we find a change in the eigendielectric polarization orientation direction with temperature which leads to the possibility that the eigenpolarization directions of the direct band-to-band transitions may be controllable by strain on the crystal lattice or by changes in internal electric fields. This approach could be utilized for design of devices operating at elevated temperatures.

We thank K. Goto and A. Kuramata and the Tamura Corp. for providing the samples studied in this investigation. This work was supported by the National Science Foundation (NSF) through the Center for Nanohybrid Functional Materials (EPS-1004094), the Nebraska Materials Research Science and Engineering Center (DMR-1420645), the Swedish Research Council (VR2013-5580), and the Swedish Foundation for Strategic Research (SSF, FFL12-0181 and RIF14-055). Partial financial support from NSF (CMMI 1337856, EAR 1521428), and J.~A.~Woollam Foundation is also acknowledged.

\bibliography{CompleteLibrary_AM}

\end{document}